\documentclass[twocolumn,preprintnumbers,
endnote,nofootinbib,amsmath]{revtex4}
\usepackage{graphicx}
\usepackage{mathtools}
\usepackage[hypertex]{hyperref}
\usepackage{subfigure}
\newcommand{\vev}[1]{\langle {#1} \rangle}
\newcommand{\lsim}{\lesssim}
\newcommand{\gsim}{\gtrsim}

\newcommand{\mP}{\bar M_P}
\newcommand{\ord}[1]{\mathcal{O}{(#1)}}
\newcommand{\beq}{\begin{equation}}
\newcommand{\eeq}{\end{equation}}
\newcommand{\bea}{\begin{eqnarray}}
\newcommand{\eea}{\end{eqnarray}}

\newcommand{\vphi}{\varphi}

\newcommand{\mhpm}{m_{H^\pm}}
\newcommand{\eq}[1]{Eq.~(\ref{#1})}

\begin{document}

\pagestyle{plain}

\title{\boldmath Right-Handed Neutrinos as the Origin of the Electroweak Scale}

\author{Hooman Davoudiasl
}
\affiliation{Department of Physics, Brookhaven National Laboratory,
Upton, NY 11973, USA}

\author{Ian M. Lewis
}
\affiliation{Department of Physics, Brookhaven National Laboratory,
Upton, NY 11973, USA}


\begin{abstract}

The insular nature of the Standard Model may be explained if the Higgs mass parameter is only sensitive to quantum corrections from physical states. Starting from a scale-free electroweak sector at tree-level, we postulate that quantum effects of heavy right-handed neutrinos induce a mass term for a scalar weak doublet that contains the dark matter particle. In turn, below the scale of heavy neutrinos, the dark matter sector sets the scale of the Higgs potential.  We show that this framework can lead to a Higgs mass that respects {\it physical naturalness}, while also providing a viable scalar dark matter candidate, realistic light neutrino masses, and the baryon asymmetry of the Universe via thermal leptogenesis.  The proposed scenario can remain perturbative and stable up to the Planck scale, thereby accommodating simple extensions to include a high scale ($\sim 2\times 10^{16}$~GeV) inflationary sector, implied by recent measurements.  In that case, our model typically predicts that the dark matter scalar is close to 1~TeV in mass and could be accessible in near future direct detection experiments.

\end{abstract}
\maketitle

\section{Introduction}

The discovery of a Higgs scalar $H$ of mass $m_H \simeq 126$~GeV at the LHC \cite{Aad:2012tfa,Chatrchyan:2012ufa}
seems to complete the Standard Model (SM).  Although the SM has been very successful,
there are strong indications that extensions of it are necessary to explain the known Universe.
Setting gravity aside, there is convincing experimental evidence for light
neutrino masses and dark matter (DM), both of which require new physics.
The SM also does not account for the
baryon asymmetry of the Universe at the observed level.

From a theoretical point
of view, the SM Higgs mechanism gives rise to a conceptual puzzle regarding the stability
of the electroweak scale, set by the vacuum expectation value (vev) of the Higgs
$\vev{H}=v/\sqrt{2}\simeq 174$~GeV, against large quantum corrections.
For example, the Yukawa coupling of the Higgs to the top quark $y_t\simeq 1$
is generally assumed to give a contribution $\sim 3 y_t^2 \Lambda^2/(8 \pi^2)$ to the Higgs potential, where
$\Lambda$ is the cutoff scale for divergent loop integrals.  The problem
arises when one identifies $\Lambda$ with the threshold for new physics, which is
often constrained to be well above the weak scale, and
perhaps as high as the reduced Planck mass ${\bar M_P}\simeq 2 \times 10^{18}$~GeV.
One then faces the question of why the Higgs mass remains near its measured value, given
the presumed quadratic sensitivity to the cutoff $\Lambda$.  This
is the well-known ``hierarchy problem."

One may assume that the hierarchy is simply removed by an appropriate fine-tuning
of various large quantum corrections against bare parameters.  However, this approach, while
consistent from a mathematical viewpoint, involves a significant amount of fine-tuning.  
A resolution of the hierarchy, generally considered more palatable, 
is to introduce new physics at a scale $\Lambda\lesssim1$~TeV that 
cures the SM of quadratic divergences.   However, the LHC data have so far offered no
evidence for any new physics up to scales $\Lambda\gsim 1$~TeV.  It thus seems at the moment
that the Higgs potential is fine-tuned.

While it is perhaps too early to draw firm conclusions about the
hierarchy problem, the lack of direct or indirect evidence for new weak scale physics has
led some to question the above assumptions.  For example, the scale $\Lambda$ is typically associated with a cutoff 
regulator and may be considered unphysical.  It has been proposed that instead of using this cutoff,
 hierarchies should be defined using the masses of physical states that interact with the Higgs.  In this view, the top Yukawa coupling
makes a contribution of $\sim y^2_t m^2_t/(8\pi^2)$ to the Higgs mass.  Since $m_t$ is at the electroweak scale, the Higgs mass is stable against this correction.  In order to to avoid fine-tuning, physical states with masses, $M$, far above the electroweak scale should have small couplings, $\lambda$, to the Higgs such that $\delta m_H^2\sim\lambda^2 M^2/(8\pi^2)\lesssim v^2$ \cite{Farina:2013mla}.  One can go further and
require that the original Lagrangian is classically scale-free and all masses are generated through
quantum effects.  We will refer to this view as the
{\it Physical Naturalness Principle} (PNP) \cite{Bardeen:1995kv,Heikinheimo:2013fta} in what follows.  (For
other related works, see also Refs.~\cite{related,Hambye:2007vf,Meissner:2007xv}.)

The assumption of PNP can potentially explain the insular nature of the SM, by removing the need
for new weak scale states with $\ord{1}$ Yukawa and gauge couplings required
to mitigate cutoff scale contributions to the Higgs mass.
 However, the PNP disfavors some popular ideas for ultraviolet (UV) physics, like grand unification and thermal
leptogenesis \cite{Farina:2013mla}.  We will focus on thermal leptogenesis, as it explains the observed
baryon asymmetry of the Universe and is intimately related to the seesaw mechanism
for light neutrino masses.

To see the problem, consider a heavy right-handed neutrino
$N$ of mass $M_N$ coupled to the Higgs via $y_N H^* {\bar L} N$, where $y_N$ is the Yukawa coupling and
$L$ is a lepton doublet in the SM.  In typical thermal leptogenesis scenarios, the
loop-induced lepton asymmetry from CP violation
is given by $\varepsilon \sim y_N^2/(8\pi)$.  Observational evidence requires $n_B/s\sim 10^{-10}$, where $n_B$ is the
baryon asymmetry and $s$ is the entropy density in the early Universe.  One can then estimate
\beq
\frac{n_B}{s} \sim \frac{\varepsilon}{g_*}\sim \frac{y_N^2}{8 \pi g_*}\,,
\label{nBovers}
\eeq
where $g_*\sim 100$ is the relativistic
degrees of freedom during electroweak phase transition.
Hence, we see that to have a viable leptogenesis mechanism, we need
\beq
y_N\gsim 5\times 10^{-4} \quad\quad {\rm (leptogenesis)},
\label{leptoyN}
\eeq
without further assumptions, such as a mass-degenerate right-handed neutrino sector \cite{Pilaftsis:1997jf}.  

The seesaw mechanism for neutrino masses $m_\nu$ gives
\beq
m_\nu \sim \frac{y_N^2 \vev{H}^2}{M_N}\,.
\label{mnu}
\eeq
For a neutrino mass of $0.1$~eV and $y_N\gtrsim 5\times 10^{-4}$, we find
$M_N \gtrsim 10^8$~{\rm GeV}.  With these constraints, the 1-loop contribution from $N$ to the Higgs mass is given by
\beq
\delta m^2_H \sim \frac{y_N^2}{4 \pi^2} M_N^2\,\gtrsim (8~{\rm TeV})^2,
\label{delm2}
\eeq
which is in conflict with PNP~\cite{OtherNeutrino}.

The conflict between the requirements for conventional leptogenesis and PNP originates from
the assumption that the particles that mediate leptogenesis are also responsible for the
seesaw mechanism.  That is, the same parameters govern leptogenesis, neutrino masses, and Higgs mass corrections.  Therefore, it seems that to avoid this situation we must
assume separate sets of fields for each scenario.  A simple solution may then be to assume that
the heavy right-handed neutrinos couple to another
scalar doublet $H_2$ and not the SM Higgs.  This can be accomplished by assuming the heavy neutrinos and $H_2$ are odd under a $Z_2$ while the SM fields are even.  Leptogenesis then proceeds through CP violating decays $N\rightarrow L H_2$, decoupling it from the SM Higgs. The $NLH_2$ Yukawa coupling must then satisfy (\ref{leptoyN}).  If $H_2$ does not get a vev it cannot be responsible for neutrino masses and will not be subject to the seesaw
constraint.

It is also interesting to
see if the introduction of $H_2$ (often referred to as an ``inert" doublet \cite{Deshpande:1977rw}) can
be motivated in other ways.  In fact,
if the extra doublet does not get a vev as suggested above, the $Z_2$ parity remains unbroken, potentially
leading to a stable DM candidate.
Note that this setup decouples the right-handed neutrinos from the SM Higgs at tree level.  Hence, it
seems that some $Z_2$-even right-handed neutrinos are needed in order to have a seesaw
mechanism for $m_\nu$.  Interestingly, it turns out that a 1-loop process can provide a
seesaw operator~\cite{Ma:2006km} and realistic $m_\nu \neq 0$, without introducing $Z_2$-even
right-handed neutrinos. Additionally, since the SM Higgs does not have direct couplings to the heavy neutrinos, the Higgs mass corrections are only sensitive to the heavy neutrino mass scale via two-loop processes, alleviating the PNP constraint.

It then appears that the above simple setup can
comply with PNP, while also accounting for the baryon asymmetry and DM content of the Universe,
as well as a mechanism for neutrino mass generation.
In what follows, we assume that the tree-level electroweak Lagrangian
has no mass scales other than $M_N$.  All other masses are generated at the loop level.  Remarkably, we will find that
a realistic Higgs potential and a good DM candidate can be achieved in this scenario,
without the need for large couplings (strong interactions) near the weak scale. In fact,
as we will illustrate, the resulting framework can address the above open questions of physics, while
remaining perturbative and stable up to the Planck scale ${\bar M_P}$.  An interesting outcome of
our framework is that masses of all fundamental particles become linearly dependent on the
right-handed neutrino mass scale $M_N$.
  In particular, while light neutrino masses $m_\nu$
arise from an effective seesaw at the weak scale, in the UV description $m_\nu$ is radiatively generated and
proportional to $M_N$.  We will next introduce a minimal model to
realize the above scenario.


\section{The Model}
\label{Model.sec}
We assume that the only massive states in this limit are
Majorana neutrinos $N_a$, $a=1,2$, with masses
$M_{N_a}$, that are odd under a $Z_2$ parity.  
There are also two scalar doublets, $H_1$ and $H_2$, 
that have the $SU(2)_L\times U(1)_Y$ quantum numbers of the
SM Higgs doublet.  We will assign a negative $Z_2$ parity to $H_2$. 

Note that $M_{N_a}$, being associated with fermions, do not give rise to a hierarchy problem.  
Nonetheless, to keep our treatment consistent, we should explain how 
the requisite Majorana masses arise from a classically scale-invariant Lagrangian.   
An interesting possibility, which we will present in the appendix, is to induce such a mass 
by the non-trivial dynamics of an asymptotically free gauge interaction \cite{Carone:1993xc}, which is scale-free at the classical level.

The mass terms and Yukawa couplings of $N_a$ are given by
\beq
-{\cal L}_N =  y^{ai} H_2^* \overline{L_i} N_a
+\frac{1}{2} M_{N_a} \overline{N^c_a} N_a +  \text{\small H.C.}\,,
\label{LN}
\eeq
where $i=1,2,3$ is the lepton generation index.
The tree-level scalar potential at high scales $\mu \gsim M_{N_i}$ has the form
\begin{eqnarray}
V_0 &=& \frac{\lambda_1}{2} |H_1|^4 + \frac{\lambda_2}{2} |H_2|^4 +
\lambda_3 |H_1|^2|H_2|^2 \nonumber \\
&+& \lambda_4 |H_1^\dagger H_2|^2 + \frac{\lambda_5}{2} \left[(H_1^\dagger H_2)^2 + \text{\small H.C.}\right]\,,
\label{V0}
\end{eqnarray}
where all coefficients are assumed to be positive.
The presence of interactions from the SM other than the Higgs potential, as well as
the requisite kinetic terms are implicitly assumed.  For simplicity, we set $\lambda_4=0$.  This coupling is responsible for splitting the charged and neutral components of $H_2$.  Hence, any isospin violating couplings to $H_2$ will generate $\lambda_4$ at loop level.  However, this will occur at the level $\sim g^4/(16\pi^2)$ for a generic coupling $g$.  For small couplings, these loops can be safely ignored and our condition $\lambda_4=0$ is maintained to a
good approximation.

At tree-level, our scalar potential [\eq{V0}] contains no mass scales and cannot lead to electroweak symmetry breaking.  However, quantum corrections can change this, as we will show via the one-loop Coleman-Weinberg potential.  In accordance with the scaleless tree-level potential, this computation is performed using dimensional regulation.  We start from the high scale and proceed with the computation in two stages.

\subsection{Scalar Masses}
To compute the Coleman-Weinberg potential, we must consider the Higgs-dependent mass matrices.  To show the key physics, let us assume that the Yukawa couplings are diagonal $y^{ia}=y_{i} \delta_{ia}$.  In this limit and writing $H_2=(H^+,\, (S+i A)/\sqrt{2})^{\rm T}$, the Lagrangian can be written as:
\begin{eqnarray}
-\mathcal{L}_N &=& \frac{1}{2} \left(\overline{\nu_i}\,\overline{N^c_i}\,\overline{\ell_i}\right)\begin{pmatrix} 0& y_i \frac{S-i A}{\sqrt{2}} & 0\\ y_{i}\frac{S-i A}{\sqrt{2}} & M_{N_i} & -y_{i} H^-\\ 0& -y_i H^-&0 \end{pmatrix} \begin{pmatrix} \nu^c_i \\ N_i \\ \ell^c_i\end{pmatrix}\nonumber \\
& +&\rm{H.C.}\,.
\end{eqnarray}
The Higgs-dependent mass eigenvalues are then zero mass states, light states
\beq
m^2_\alpha(H_2) = \frac{M_{N_\alpha}^2}{2}\left(1+2 y_\alpha^2\frac{|H_2|^2}{M_{N_\alpha}^2}-\sqrt{1+4 y_\alpha^2\frac{|H_2|^2}{M_{N_\alpha}^2}}\right)
\eeq
and heavy states
\beq
M^2_{\alpha}(H_2)= \frac{M_{N_\alpha}^2}{2}\left(1+2 y_\alpha^2\frac{|H_2|^2}{M_{N_\alpha}^2}+\sqrt{1+4 y_\alpha^2\frac{|H_2|^2}{M_{N_\alpha}^2}}\right),
\eeq
where $\alpha=1,2$.
The contribution from the above states to the effective potential can be obtained from
\begin{eqnarray}
V_1(H_2, \mu) &=& -\frac{1}{32\pi^2} \sum_{\alpha=1}^2\bigg\{ M_\alpha^4(H_2)\label{V1H2sum}\\
&\times&\left[\log\left(\frac{M_\alpha^2(H_2)}{\mu^2}\right) - \kappa_N-\frac{1}{2}\right]\nonumber\\
&+& m_\alpha^4(H_2)
\left[\log\left(\frac{m_\alpha^2(H_2)}{\mu^2}\right) - \kappa_N-\frac{1}{2}\right]\bigg\}\nonumber
\end{eqnarray}
where $\mu$ is the renormalization scale. The constant $\kappa_N$ has been introduced to parameterize the renormalization scheme dependence of the effective potential with $\kappa_N=1$ corresponding to the $\overline{\rm MS}$ scheme. A straightforward calculation yields
\begin{eqnarray}
V_1(H_2, \mu) &=& \sum_\alpha \frac{y_{\alpha}^2 M_{N_\alpha}^2}{8 \pi^2}
\left[\kappa_N - \log\left(\frac{M_{N_\alpha}^2}{\mu^2}\right)\right] |H_2|^2\nonumber \\
&+&  \ldots\,,\label{V1Himu}
\end{eqnarray}
Since there are no other mass scales in the Lagrangian, this is the only contribution to the scalar mass parameters.  As can be clearly seen, for $\kappa_N$ not too large, as may be expected for a perturbative quantity, the mass of $H_2$ is loop suppressed compared to $M_N$. Hence, to determine the structure of the theory at the DM scale it is more appropriate to work in an effective field theory (EFT) in which the neutrinos are integrated out.  This is necessary since for $\mu\ll M_N$, the log in Eq.~(\ref{V1Himu}) becomes large and we will see the EFT approach can alleviate this potential issue.

Integrating out the heavy neutrinos is accomplished via matching the high energy scalar potential to an effective potential valid at scales below the neutrino mass $M_N$.  For now, we neglect other effects of integrating out the heavy neutrinos. We will return to this subject and how it relates to light neutrino masses, in the next section.  In the EFT below $M_N$, the induced $H_2$ mass is accounted for by introducing a ``tree-level" mass, $\mu_2$, for $H_2$:
\begin{eqnarray}
V_0 \rightarrow V_0+\mu^2_2 |H_2|^2
\label{V0med}
\end{eqnarray}
The contribution of $H_2$ to the one-loop Coleman-Weinberg potential is~\cite{Coleman:1973jx}
\begin{eqnarray}
V_1(H_2, H_1, \mu)&=&-\frac{\mu^2_2}{16\pi^2}\left(\kappa_2-\log\frac{\mu^2_2}{\mu^2}\right)\label{V1med}\\
&&\times\left(3 \lambda_2 |H_2|^2+2 \lambda_3 |H_1|^2\right)+\ldots,\nonumber
\end{eqnarray}
where we have introduced a second renormalization scheme constant $\kappa_2$ that is in principle different from $\kappa_N$.  Again, $\kappa_2=1$ corresponds to the $\overline{\rm MS}$ scheme.
We will work under the simplifying assumption $M_{N_1}=M_{N_2}=M_N$ and $y_1=y_2=y_N$.  Matching the two potentials at a scale $\mu=M_N$, we obtain a mass parameter\footnote{Although the matching scale may not be precisely $M_N$, any variation in the scale can be absorbed into the $\kappa$s.  In the numerical results, variation in $\kappa$ encompasses the renormalization scheme and matching scale dependence of our results}:
\beq
\mu^2_2=\frac{M_N^2 y^2_N\kappa_N}{4\pi^2}\left[1+\frac{3\lambda_2}{16\pi^2}\left(\kappa_2-\log\frac{y^2_N\kappa_N}{4\pi^2}\right)\right].
\label{mu2}
\eeq
While the second term of Eq.~(\ref{mu2}) is beyond one-loop order in the parameters of the high energy theory, for now we keep it for illustrative purposes.    Since we require a DM candidate from $H_2$, the original $Z_2$ must remain unbroken.  We need $\mu^2_2>0$ and
hence $\kappa_N>0$. We also note that the potentially destabilizing large log of the high energy theory in Eq.~(\ref{V1Himu}) has also been replaced by a loop suppressed log in the EFT.

Adherence to the PNP implies
\beq
\kappa\sim\mathcal{O}(1),
\eeq
for any general $\kappa$-scheme.  If $\kappa\ll1$, then the loop effects of the physical mass scale are fine-tuned against a counter-term, violating PNP.  In addition, if $\kappa\gg1$,  the counterterm must be much larger than the loop effects.  This is numerically similar to adding a tree-level Higgs mass to the Lagrangian.

From Eq.~(\ref{V1med}), we see that $H_2$ induces a loop-suppressed $H_1$ mass parameter.  Hence, similar to the above procedures, we integrate out $H_2$ and match onto the $H_1$ potential valid for $\mu<\mu_2$:
\beq
V_0 =-\mu^2_1 |H_1|^2+ \frac{\lambda_1}{2} |H_1|^4,
\label{V0low}
\eeq
where in anticipation of the result we introduce a tachyonic mass for $H_1$.
Matching the potentials at a scale of $\mu=\mu_2$, we find the mass parameter for $H_1$ to be
\beq
\mu^2_1 = \frac{\lambda_3\kappa_2}{8\pi^2}\mu^2_2\left[1+\frac{3\lambda_1}{16\pi^2}\left(\kappa_1-\log\frac{-\lambda_3\kappa_2}{8\pi^2}\right)\right],
\label{mu1mu2}
\eeq
where we have again introduced another $\kappa_1\sim\mathcal{O}(1)$.
There are a few interesting things to note about this result.  If $\kappa_2\lambda_3>0$ the mass of $H_1$ is tachyonic and leads to electroweak symmetry breaking (See also Ref.~\cite{Hambye:2007vf}.) Additionally, it is interesting to note that $\mu_1$ does not depend $\lambda_5$.  This can be understood by noting that the mass term is of the form $H_1^\dagger H_1$, while, due to the structure of the coupling, $\lambda_5$ could only contribute to terms like $H_1 H_1$, which are forbidden by electroweak symmetry.  Finally, for $\kappa_2\lambda_3>0$, Eq.~(\ref{mu1mu2}) contains $\log(-1)$, apparently indicating an imaginary potential.  This is an artifact of expanding the effective potential around $H_1=0$, which is not the true vacuum for positive $\kappa_2\lambda_3$.  If the potential is expanded around $\vev{H_1}=v/\sqrt{2}$ the $\log(-1)$ does not appear.

Interestingly, if we use the ${\rm MS}$ scheme where only the $1/\epsilon$ poles are cancelled, positive $\kappa_1,\kappa_2,$ and $\kappa_N$ are obtained.  That is, the finite part of the one-loop correction generates a positive $\mu_1^2$ and $\mu_2^2$.  If scalar mass parameters are loop generated, we may expect the finite pieces to be the dominant effect and any counterterm may be required to be subdominant.  In that case, the above scenario naturally leads to the correct symmetry breaking pattern. 

Since we expect both $\kappa_N$ and $\kappa_2$ to be order one (and positive, from physical considerations in our model), for simplicity we set
\beq
\kappa_2=\kappa_N\equiv\kappa.
\eeq
Then in the numerical results, variation of $\kappa$ will encompass our renormalization scheme and matching scale uncertainty\footnote{We could have chosen to work in the $\overline{\rm MS}$ scheme with $\kappa_N=\kappa_2=\kappa_1=1$.  However, it is not then clear that the masses Eqs.~(\ref{mu2}) and (\ref{mu1mu2}) are the same as those physical pole-masses and couplings we wish to know for DM and collider searches. The $\kappa$s could be determined if all the pole masses were measured.  However, this is obviously not the case yet for DM or heavy neutrinos.}.   To obtain the correct symmetry breaking pattern, we then need $\lambda_3>0$.  Putting all the above results together and dropping higher order terms, we finally obtain
\beq
\mu^2_1\approx \frac{\lambda_3 y^2_N}{32\pi^4}M^2_N{\kappa}^2.
\label{mu1Neu}
\eeq
Hence, $\mu_1^2$ is suppressed by two-loops compared to $M^2_N$, alleviating the PNP constraint.

In order to obtain the SM Higgs boson mass $m_H \simeq 126$~GeV and $v=246$~GeV, we need $\mu_1^2 \simeq (89~{\rm GeV})^2$.  Using the leptogenesis requirement (\ref{leptoyN}) and Eq.~(\ref{mu1Neu}), we find the heavy neutrino mass is constrained to be
\beq
M_{N}\lesssim \frac{5\times 10^4~{\rm TeV}}{\sqrt{\lambda_3}\,\kappa\,(y_N/10^{-3})}.
\label{MN1PNP}
\eeq
Although these values satisfy the requirement of leptogenesis and PNP, we must now determine if they can generate 
light neutrino masses $\sim 0.1$~eV required to explain neutrino oscillation data.

\subsection{Light Neutrino Masses}
\begin{figure}[tb]
\centering
\includegraphics[width=0.2\textwidth,clip]{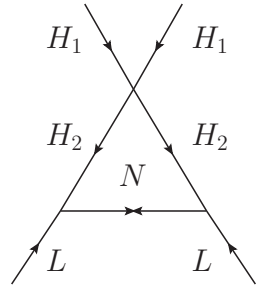}
\caption{One loop diagram contributing to neutrino mass.}
\label{numass}
\end{figure}
Since the heavy neutrinos do not couple to the electroweak symmetry breaking Higgs field at tree level in this model,
the above setup does not allow for the conventional seesaw mechanism.  However,
the tree level couplings of $H_2$ to $N_{1,2}$ and $H_1$ allow for a 1-loop realization
of the seesaw mechanism which yields light (SM) 
neutrino masses \cite{Ma:2006km,Hambye:2009pw,Chen:2009gd}, 
as shown in Fig.~\ref{numass}.  For $M_N\gg \mu_2$, the relevant operator is
\begin{eqnarray}
\mathcal{L}_{\rm eff}&=&-\sum_\alpha \frac{y_\alpha^2 \lambda_5}{16\pi^2M_{N_\alpha}}\left(1+\log\frac{\mu_2^2}{M^2_{N_\alpha}}\right) 
H_1 \overline{L^c} H_1 L\nonumber\\
&+& {\rm H.C.}\, .
\end{eqnarray}
Using the previous results, this gives a neutrino mass
\begin{eqnarray}
m_\nu &\approx& -\frac{\lambda_5 \, y_N^2 \, v^2}{8 \pi^2\, M_N}
\left[\log\left(\frac{4\pi^2}{y_N^2 \kappa}\right)-1\right].\nonumber
\label{1-loop-mnu}
\end{eqnarray}
From the leptogenesis condition (\ref{leptoyN}), PNP condition (\ref{MN1PNP}), and setting $m_\nu=0.1$~eV, we find that
\begin{eqnarray}
|\lambda_5|&\lesssim& \frac{0.3
}{\sqrt{\lambda_3}\kappa}\approx\frac{\mu_2}{\sqrt{\kappa}~2.7~{\rm TeV}},
\label{lam5upper}
\end{eqnarray}
where we have dropped a small $\log\kappa$.
Hence, for reasonable values of $\lambda_5$ and $\lambda_3$ this scenario can accommodate leptogenesis, neutrino mass, and PNP.

\section{Dark Matter Candidate}
Note that since $\vev{H_2}=0$ the above construct leaves the $Z_2$ symmetry intact. Hence, the lightest parity-odd particle is stable. One can easily check that the with $\vev{H_1} = v/\sqrt{2}$, the masses of the scalar states are given by
\bea
m_h^2 &=& \lambda_1 \,v^2 \\ \nonumber
m_S^2 &=& \mu_2^2 + \lambda_S \,v^2 \\ \nonumber
m_A^2 &=& \mu_2^2 + \lambda_A \,v^2 \\ \nonumber
\mhpm^2 &=& \mu_2^2 + \frac{\lambda_3}{2}\, v^2\,,
\label{scalarmasses}
\eea
where $\lambda_S\equiv (\lambda_3+\lambda_4+\lambda_5)/2$ and $\lambda_A\equiv (\lambda_3+\lambda_4-\lambda_5)/2$.  To avoid having a stable charged particle, we need to make sure the charged state is the not the lightest. A simple choice that would satisfy this is $\lambda_4=0$, making $A$ the lightest $Z_2$ odd particle and hence a DM candidate, which we will assume for purposes of illustration in what follows.

\begin{figure*}[tb]
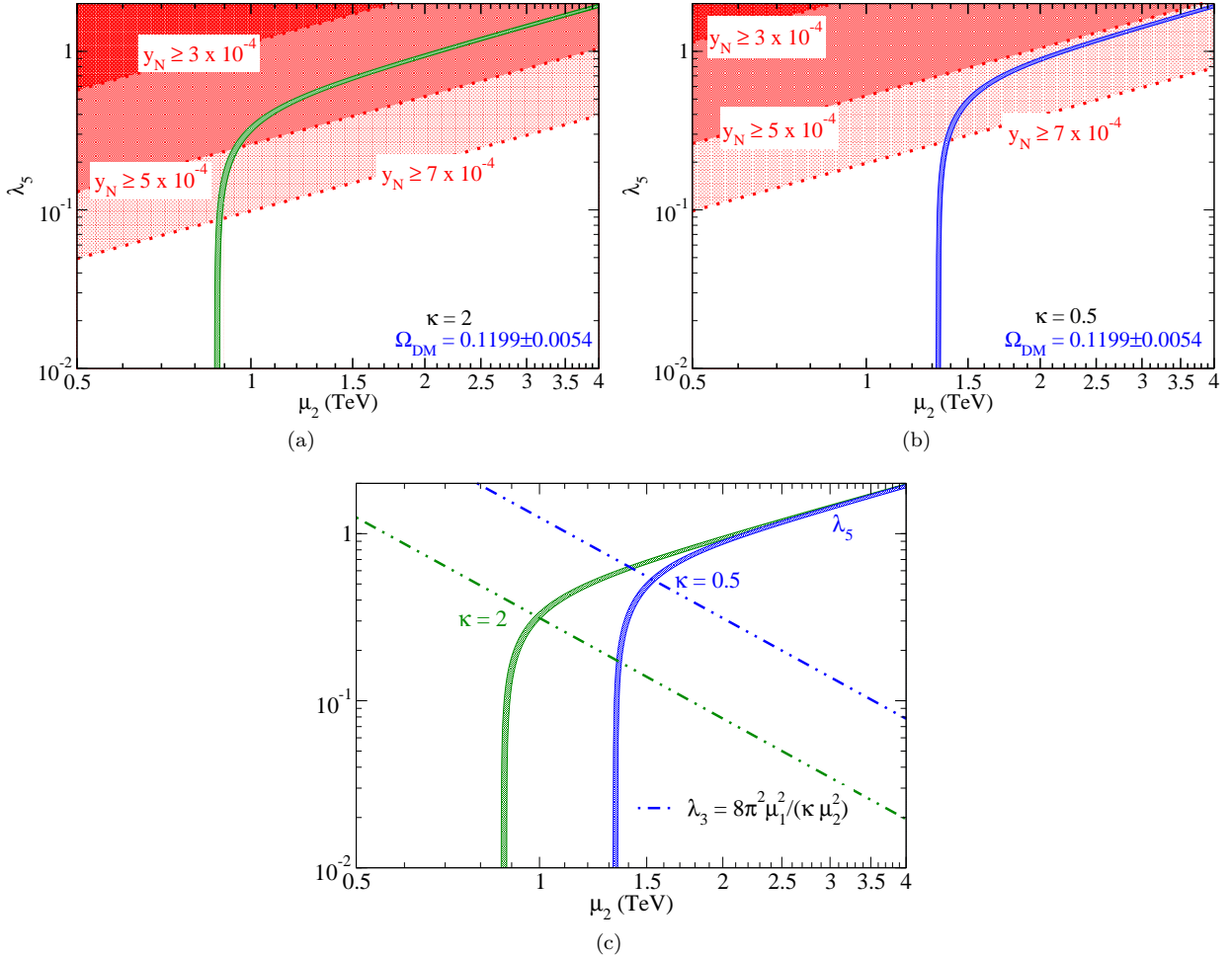

\centering
\subfigure[]{
\includegraphics[width=0.45\textwidth,clip]{DMlam5lam3kap2.eps}\label{DMkap5}}
\subfigure[]{
\includegraphics[width=0.45\textwidth,clip]{DMlam5lam3kap0p5.eps}\label{DMkap05}}\\
\subfigure[]{
\includegraphics[width=0.45\textwidth,clip]{DMlam5kap.eps}\label{DMboth}}
\caption{Values of (a,b,c) $\lambda_5$ (green and blue shaded) and (c) $\lambda_3$ (dash-dot-dot) as required by being within $2\sigma$ of the DM constraint in Eq.~(\ref{Planck}) and mass relationship of Eq.(\ref{mu1mu2}), respectively. The results are shown for (a) $\kappa=2$, (b) $\kappa=0.5$, and (c) both $\kappa=2$ and $0.5$.  In (a) and (b), the red dotted lines are the upper bounds on $\lambda_5$ values that obey PNP and neutrino mass constraints for leptogenesis bounds of $y_N\gtrsim3\times10^{-4}$, $5\times10^{-4}$, and $7\times10^{-4}$.  }
\label{DM}
\end{figure*}
With the above choice of parameters, we have $m_S>\mhpm>m_A$.  The mass splitting between the charged and neutral states, according to \eq{scalarmasses}, is given by
\beq
\Delta \approx \frac{\lambda_5 \,v^2}{4\mu_2}.
\label{Delta}
\eeq
Assuming $\Delta\ll M_W$, a rough order of magnitude estimate of the decay rate governed by such a mass splitting can be obtained
from
\beq
\Gamma_\Delta \sim \frac{G_F^2}{64 \pi^3} \Delta^5\,,
\label{GamDel}
\eeq
where $G_F$ is Fermi's constant.  The unstable states, $H^\pm$ and $S$, should decay before Big Bang Nucleosynthesis (BBN), at about $t\sim 1$~s.  With this requirement we find $\Delta\gtrsim 5$~MeV.  This translates into a lower limit on $\lambda_5$:
\beq
\frac{\mu_2}{2.7\times10^3~{\rm TeV}}\lesssim |\lambda_5|.
\label{lam5lower}
\eeq
As can be seen, the requirement that the unstable states decay before BBN is compatible with our previous upper limit in Eq.~(\ref{lam5upper}) and is not a stringent constraint on $\lambda_5$.  After electroweak symmetry breaking, there are also electroweak corrections coming from the SM $W^\pm$ and $Z$ that would further raise the mass of $H^\pm$ {\it above} the neutral states by $\ord{10}$~MeV, which further reduces any unwanted effects from $H^\pm$ decays by making their rate larger.  However, these corrections do not split the neutral components of the inert doublet.

Considering only the BBN  constraints, the neutral states could be completely degenerate ($\lambda_5=0$) and the mass splitting between the charged and neutral states due to electroweak corrections would be sufficient to guarantee a fast enough decay of $H^\pm$.  However, if the neutral states are degenerate, then DM could scatter in direct detection experiments via $Z$ boson exchange, 
which would be in severe conflict with experimental bounds.  This constraint can be alleviated by noting that with a splitting between the neutral states, DM direct detection through $Z$ exchange would require an inelastic up-scattering to a state that is heavier by $\Delta$ \cite{Barbieri:2006dq}.  Hence a non-zero $\lambda_5$ is needed to avoid direct detection experiments.  The typical kinetic energy of a TeV-scale DM particle, corresponding to a virial velocity of order 200~km/s, is $\sim {\rm few}\times 100$~keV, which is small compared to $\Delta\gtrsim 5$~MeV as required by BBN.  Hence, detection through $Z$ exchange is well-suppressed and does not pose a phenomenological
constraint.  DM scattering from nucleons through Higgs exchange is still possible \cite{DMHiggs},
due to the coupling proportional to $\lambda_A$, with a cross section \cite{Hambye:2009pw}
\beq
\sigma_n \simeq \frac{f_n^2\,\lambda_A^2}{\pi}\frac{m_n^4}{\mu_2^2\, m_H^4},
\label{sigman}
\eeq
where $f_N\simeq 0.3$ and $m_n\simeq 1$~GeV.
However, for DM masses of order $\mu_2\sim 1$~TeV and $\lambda_5 \sim 0.01-1$, one gets $\sigma_n \sim 1-5\times10^{-45}$~cm$^2$,
which is consistent with current limits from the LUX experiment \cite{Akerib:2013tjd},
but could be within the reach of near future direct detection measurements.

We calculate the relic density of the dark matter particle $A$ using the thermally averaged cross section in Ref.~\cite{Hambye:2009pw} and give the resulting formulas under our assumptions in the appendix.  With $\lambda_4=0$, this calculation depends on the couplings $\lambda_3$ and $\lambda_5$ and the mass $\mu_2$, in addition to the electroweak gauge couplings\footnote{We have checked the consequences of including $\lambda_4$ in our calculations and our conclusions are not changed significantly.}.  The requirement that we reproduce the correct SM Higgs mass and vev gives a relationship between $\lambda_3$ and $\mu_2$ via Eq.~(\ref{mu1mu2}).  Hence, for the DM calculation there are only two free parameters, which are chosen to be $\lambda_5$ and $\mu_2$.  We require that the DM relic density is within $2 \sigma$ of the current Planck results~\cite{Ade:2013zuv}:
\beq
\Omega_{DM} h^2=0.1199\pm0.0027.
\label{Planck}
\eeq

The shaded blue and green bands in Fig.~\ref{DM} show the allowed values for $\lambda_5$ and $\mu_2$ that obey the relic density constraint within $2\sigma$.  Figure~\ref{DM} shows the result for (a) $\kappa=2$, (b) $\kappa=0.5$, and (c) both $\kappa=2$ and $0.5$.  For comparison purposes, in Fig.~\ref{DMboth} we also include the results for $\lambda_3$ from Eq.~(\ref{mu1mu2}). If $\mu_2\gtrsim 550$~GeV, the annihilation purely from gauge interactions is insufficient to reproduce the observed abundance, and would overclose the Universe~\cite{Hambye:2009pw}.  Hence, co-annihilation via the scalar quartic terms are essential to obtaining the correct relic abundance and there is a lower bound on their combined contribution to the thermally averaged cross section.  For $\lambda_5\ll\lambda_3$, the coupling $\lambda_5$ can be neglected and the relic density constraint fully determines $\mu_2\approx 0.8$~TeV ($1.3$~TeV) for $\kappa=2$ ($0.5$).  These values correspond to lower bounds on the scale of DM.   In the limit $\mu_2\gg 1$~TeV, $\lambda_3$ can be neglected and we obtain the equality
\beq
\lambda_5\approx \frac{\mu_2}{2~{\rm TeV}},
\label{lam5approx}
\eeq
independent of $\kappa$, as evident from Fig.~\ref{DMboth}.  One interesting consequence of this equation is that since $\lambda_5/\mu_2$ is fixed, we obtain a maximum value of the mass splitting $\Delta\approx 7$~GeV.  We note that for values of scalar mass splittings typical in our work, the results of Ref.~\cite{Hambye:2009pw} suggest that 
our model parameter space is not constrained by electroweak precision data.

In Figs.~\ref{DMkap5} and (b), the red dotted lines indicate the upper limits on $\lambda_5$ that are compatible with the PNP, leptogenesis, and neutrino mass.  The leptogenesis bound in Eq.~(\ref{leptoyN}) is a rough approximation.  To show the effect of order one variations we show the bounds on $\lambda_5$ using a leptogenesis bound of $y_N\gtrsim3\times10^{-4}$, $5\times10^{-4}$, and $7\times10^{-4}$.  The bound on $\lambda_5$ using $y_N\gtrsim 5\times10^{-4}$ is given in Eq.~(\ref{lam5upper}).  The regions above the dotted lines are in conflict with our requirements and are shaded red.
A more complete calculation of leptogenesis in our scenario is needed to determine the precise bound.  However, as can be clearly seen, the allowed mass scales for DM greatly depend on the value of $y_N$.

\section{Running Couplings}

\begin{figure*}
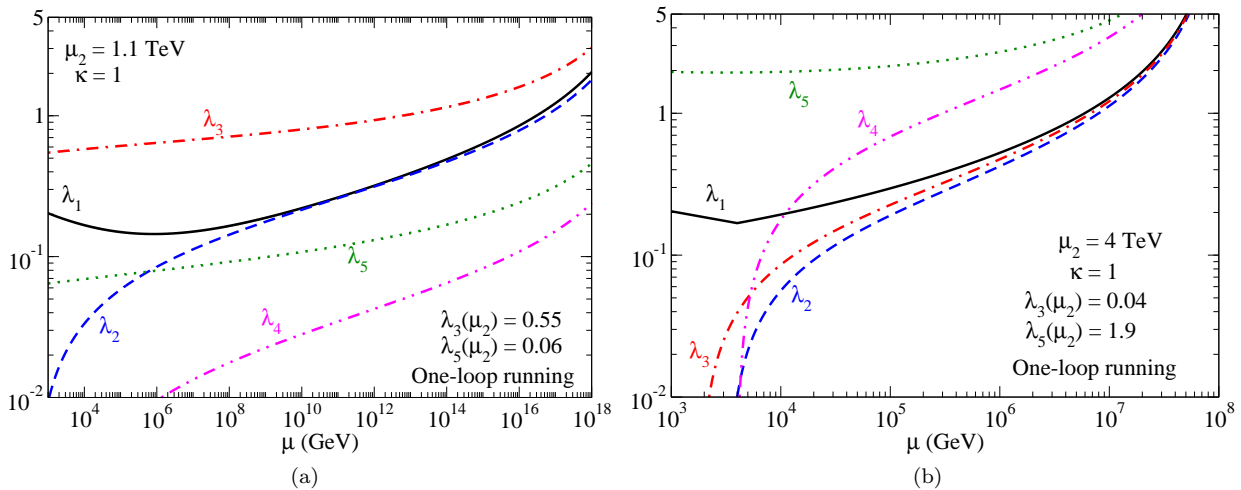

\subfigure[]{
\includegraphics[width=0.45\textwidth,clip]{PlanckRunning.eps}\label{Planckrun}}
\subfigure[]{
\includegraphics[width=0.45\textwidth,clip]{MNrunning.eps}\label{MNrun}}
\caption{Running scalar quartic couplings versus the renormalization scale $\mu$ the two points (a) $\mu_2=1.1$~TeV and (b) $\mu_2=4$~TeV  with $\kappa=1$.}
\label{running}
\end{figure*}
We now examine the perturbativity of our scalar quartic couplings at high scales.  For initial conditions we find $\lambda_1=0.26$ from SM Higgs mass and vev values at the scale $\mu=m_H$, while the other quartics are set at the scale of DM $\mu=\mu_2$.  The coupling $\lambda_3$ is fixed by Eq.~(\ref{mu1mu2}), $\lambda_5$ is fixed by the relic density constraint at a given $\mu_2$ and $\kappa$, $\lambda_4=0$, and we set $\lambda_2=0.01$.  We perform a one-loop analysis using the renormalization group (RG) equations in Ref.~\cite{Hill:1985tg}.

Figure~\ref{running} shows the results of the one-loop running as a function of the renormalization scale $\mu$.  We choose $\kappa=1$ as an illustrative value and use (a) $\mu_2=1.1$~TeV and (b) $\mu_2=4$~TeV as benchmark points.  As can be seen in Fig.~\ref{MNrun}, for the parameter region consisistent with DM, the quartic couplings remain perturbative to at least $M_N$.  In fact, near $\mu_2\approx 1.1$~TeV, the lower bound on $\mu_2$ for $\kappa=1$, the couplings stay perturbative to beyond the reduced Planck scale $\bar{M}_P\sim 10^{18}$~TeV.

If a quartic coupling obtains a Landau pole before $\bar{M}_P$, one may worry about the consistency of the approach advocated here~\cite{Meissner:2007xv}.  A Landau pole in a quartic coupling introduces a high energy scale that strongly couples to the scalars.  The scalar masses may then receive large quantum corrections and be pulled up to this scale. Hence, it is reasonable to demand that the couplings stay perturbative to $\bar{M}_P$.  In this case, we are drawn to the conclusion that DM should be very near $1$~TeV.

Additionally, the recent detection of $B$-mode polarization of cosmic microwave background \cite{Ade:2014xna}
is a possible indication for inflation at a scale of $\sim 2\times 10^{16}$ GeV.
To embed the scenario presented here in a realistic inflation model, one may expect that the couplings need to stay perturbative to the scale of inflation.  As indicated by the running, if $\mu_2$ is much above $1$~TeV, the couplings become strong before $10^{16}$~GeV.  Hence, considering inflation in addition to the previous constraints, we may expect the scale of dark matter to be quite close to $1$~TeV.

\section{Conclusions}

The smallness of the Higgs mass compared to large scales of physics is often assumed to be a puzzle whose resolution requires new physics near the
weak scale.  However, one is then faced with the experimental puzzle of why such new physics has not been found at high energy experiments or in precision measurements.
One may trace the source of this conflict to the assumption that the Higgs mass is sensitive to arbitrarily high energy scales through
$\ord{1}$ Standard Model (SM) couplings, such as the top Yukawa coupling.

As an alternative point of view, one may adopt the ``physical naturalness principle (PNP)" which postulates a scale-free
classical Lagrangian whose mass scales are generated through quantum effects.  Here, only physical masses, not arbitrary cutoff regulators, can affect the Higgs potential.  In that view, the top (or any other SM states) do not destabilize the weak scale and the effect of any high scale particles can be suppressed if they have small couplings to the Higgs.  This simple assumption is not without consequence.  For example, the right-handed neutrinos in the usual seesaw scenario cannot have sizable couplings to the Higgs (or else PNP would be violated) which seems to rule out generic leptogenesis scenarios.

In this work, we assumed the PNP and examined how to
reconcile its requirements with those of leptogenesis and a realistic seesaw mechanism for neutrino masses.  Furthermore, we
assumed that the underlying electroweak theory is classically scale invariant, and all the mass scales are generated through quantum loop effects from heavy right-handed neutrinos.  These heavy fermions are responsible for both leptogenesis and light neutrino masses.  This setup naturally leads to the assumption of an extra scalar doublet charged under a $Z_2$ parity, hence providing a dark matter candidate.  We found that this simple model can lead to viable dark matter from the extra scalar doublet, realistic neutrino masses, and successful leptogenesis, while respecting PNP.  A generic prediction of our model is that dark matter and its associated weak doublet states are nearly degenerate and characterized by a mass $\sim 1$~TeV.  These scalars may only be accessible at near future direct detection experiments or future hadron colliders operating well above the LHC center of mass energy \cite{Low:2014cba}.

We showed that the above scenario can be realized while maintaining a perturbative parameter space and stable scalar potentials, up to the Planck scale.  Hence our framework can be a natural complement to simple models of inflation that are characterized by high scales $\sim 2\times  10^{16}$~GeV, as recent cosmological measurements seem to demand.  This requirement, which may be needed for the self-consistency of the approach 
adopted in our work~\cite{Meissner:2007xv}, suggests that the dark matter mass is close to 1~TeV.  One may worry that the 
inflationary scale may introduce large quantum corrections to our scalar sector.  However, as illustrated here, this depends on how strongly the inflaton couples to the scalar sector, and PNP would indicate that this coupling should be very small.

\acknowledgments

We thank P. Meade and A. Strumia for discussions.  Work supported in part by the United States Department of
Energy under Grant Contracts DE-AC02-98CH10886.

\appendix
\section{A scenario for the origin of right-handed neutrino masses} 
Here, we outline a classically scale-invariant scenario for generating the requisite 
masses of the right-handed neutrinos, denoted here as $M_N$.  The nuetrino mass will be generated via the vev of a scalar singlet, $\vphi$.  In order to obtain $\vev{\vphi}\neq 0$, we include massless fermions $\psi_L$ and $\psi_R$, that are in the fundamental representation of an 
$SU(n)$ Yang-Mills gauge interaction.  This gauge interaction is asymptotically free and becomes confining 
at a scale $f_n > M_N$, as can be arranged by an appropriate choice of $n$ and the gauge 
coupling $g_n \lsim 1$ at $\mP$.    We can write down the following scale-free interactions 
\beq
-{\cal L}_\vphi =\frac{\lambda_\vphi}{2} \, \vphi^4 +\left(\frac{1}{2} c_N \, \vphi  \overline{N^c} N - c_{\psi}\, \vphi\,  \overline{\psi_{L}} \psi_{R} 
+  \text{\small H.C.}\right), 
\label{Lvphi}
\eeq
where $c_N, c_{\psi} >0$ are Yukawa couplings; their signs are chosen for later ease of notation.   
Here, $\lambda_\vphi$ denotes the $\vphi$ quartic 
self-coupling.  We will assume that all other 
couplings to the SM and the Higgs doublet sectors are tiny and negligible.
This Lagrangian respects the $Z_2$ parity of Section~\ref{Model.sec}.

Once the $SU(n)$ interactions become strong, we expect to have $\vev{\overline{\psi_{L}} \psi_{R}}\sim f_n^3$.  
The above couplings in (\ref{Lvphi}) then imply that $\vphi$ will develop a non-zero vev given by 
\beq
\vev{\vphi} \sim \left(\frac{c_{\psi}}{\lambda_\vphi}\right)^{1/3} f_n \, .
\label{vev-vphi}
\eeq  
The scalar $\vphi$ then has a mass 
\beq
m_\vphi^2\sim \lambda_\vphi \vev{\vphi}^2
\eeq
The above mechanism for generation of $\vev{\vphi}\neq 0$ is similar in spirit to that of 
Ref.~\cite{Carone:1993xc}.  In order to avoid having Landau poles or instabilities, it is sufficient 
to assume that $c_N, c_{\psi}, \lambda_\vphi \ll 1$ at the scale $\mu = f_n$.  
The mass of the right-handed neutrinos is given by 
\beq
M_N = c_N \vev{\vphi}\, .
\label{NRmass}
\eeq 

The gauge interactions of $\psi_{L,R}$ have a cihral symmetry $U(1)_L\times U(1)_R$, which is broken at the condensation scale $f_n$, leading to massless pion $\pi_n^0$.  However, the Yukawa term proportional 
to $c_{\psi}$ explicitly breaks the chiral symmetry and leads to non-zero pion mass:
\bea
m_{\pi_n^0}^2 \sim c_{\psi}\vev{\vphi} f_n,
\label{mpi}
\eea
Hence, for $c_N\ll c_\psi, \lambda_\vphi$  we have 
\beq
M_N\ll m_{\pi^0_n}, m_\vphi
\eeq
Let us consider $M_N \sim 10^8$~GeV, typical 
of our model, as discussed earlier.  For $c_N\sim 10^{-3},c_\psi \sim \lambda_\vphi\sim 10^{-2}$, we then have a condensation scale $f_n \sim 10^{11}$~GeV (say, for $n=4$ and $g_n\simeq 0.6$ at $\bar{M}_P$.)  We get for the pion mass $m_{\pi^{0}_n}\sim 10^{10}$~GeV and scalar mass $m_\vphi\sim10^{10}$~GeV.  Hence, for reheat temperatures in the range $10^8-10^{10}$~GeV, thermal leptogenesis is viable and the new scalar and composite states will not be present in the early universe.

\section{Thermally Averaged Cross Section}
Taking into account coannihilations between different species, the thermally averaged cross section is~\cite{Hambye:2009pw}
\begin{eqnarray}
\vev{\sigma_{eff}v}=\sum_{i,j=1}^4 \vev{\sigma^{ij}v}\frac{n^{eq}_i}{n^{eq}}\frac{n^{eq}_j}{n^{eq}},
\end{eqnarray}
where the $\{i,j=1,2,3,4\}$ refer to the scalar components $\{S,A,H^+,H^-\}$ and $\vev{\sigma^{ij}v}$ is the thermally averaged coannihilation cross section between species $i,j$.  The equilbrium number densites are given by
\beq
n^{eq}_i=\left(\frac{m_i T}{2\pi}\right)^{3/2} e^{-m_i/T}
\eeq
and $n^{eq}=\sum_i n^{eq}_i$.  As discussed above, the mass splitting, $\Delta$, between the different scalars is small compared to the overall mass scale $\mu_2$.  Hence, $n^{eq}_i\approx n^{eq}_1$  and $n^{eq} \approx 4 n^{eq}_1$ up to corrections of order $\Delta/\mu_2\ll 1$.  The thermally averaged cross section can then be simplified to
\beq
\vev{\sigma_{eff}v}\approx\frac{1}{16}\sum_{i,j=1}^4 \vev{\sigma^{ij}v}.
\eeq

From Ref.~\cite{Hambye:2009pw}, in the $s$-wave approximation the coannihilation cross sections are given by
\beq
\vev{\sigma^{ij}v}\approx A_{0}^{ij}+\frac{\Lambda^{ij}}{32\pi m^2_{S}},
\eeq
where $A_{0}^{ij}$ ($\Lambda^{ij}$) parameterize the gauge (quartic scalar) interactions.  The results for $A_{0}^{ij}$ and $\Lambda^{ij}$ are given by Eqs. (3.15) and (3.17) in the published version of Ref.~\cite{Hambye:2009pw}, respectively.
Under our assumption of $\lambda_4=0$ and using $\Delta\ll \mu_2$,
the result for the thermally averaged cross section achieves the simple form
\beq
\vev{\sigma_{eff}v}\approx\frac{1}{512\pi\mu_2^2}\left[\left(3-2 s_W^4\right)\left(\frac{g}{c_W}\right)^4+8\lambda_3^2+12\lambda_5^2\right],
\label{thermave}
\eeq
where from Eq.~(\ref{mu1mu2})
\beq
\lambda_3\approx\left(\frac{790~{\rm GeV}}{\mu_2}\right)^2\kappa^{-1}.
\eeq

In the $s$-wave approximation, the relic density is then given by
\beq
\Omega_{DM}h^2\simeq \frac{1.04\times 10^9~{\rm GeV}^{-1}x_F}{\sqrt{g_*}\,M_P\,\vev{\sigma_{eff}v}},
\eeq
where $x_F=m_A/T_F$ is set by the freeze out temperature $T_F$, $g_*\approx 100$ is the number of relativistic degrees of freedom at freeze-out, and $M_P=1.22\times10^{19}$~GeV is the Planck mass.  The freeze out temperature can be found numerically from
\beq
x_F=\ln\frac{0.038\,M_P\,g_{eff}\,m_A\,\vev{\sigma_{eff}v}}{\sqrt{g_*\,x_F}},
\eeq
where $g_{eff}=\sum_i n^{eq}_i/n_1^{eq}\approx 4$.  We find for our region of interest $x_F\approx 25$.

From the above results, we can use $\Omega_{DM}h^2=0.12$~\cite{Ade:2013zuv} and the above approximations to find values for $\mu_2$ and $\lambda_5$ in various limits.  For $\lambda_5\ll \lambda_3$, the thermally averaged cross section in Eq.~(\ref{thermave}) is completely determined by $\mu_2$.  To obtain the correct relic abundance, we find to an accuracy of a few percent
\beq
\left(\frac{\mu_2}{\rm TeV}\right)^2\approx \kappa^{-2/3}+0.1+0.01\,\kappa^{2/3}.
\eeq
Similarly, for $\mu_2\gg 1$~TeV, $\lambda_3\ll1$ can be neglected.  In this case, we find the relationship
\beq
\lambda_5\approx 0.49\sqrt{\left(\frac{\mu_2}{\rm TeV}\right)^2-0.31},
\eeq
independent of $\kappa$.


\begin{thebibliography}{99}

\bibitem{Aad:2012tfa}
  G.~Aad {\it et al.}  [ATLAS Collaboration],
  Phys.\ Lett.\ B {\bf 716}, 1 (2012)
  [arXiv:1207.7214 [hep-ex]].

\bibitem{Chatrchyan:2012ufa}
  S.~Chatrchyan {\it et al.}  [CMS Collaboration],
  Phys.\ Lett.\ B {\bf 716}, 30 (2012)
  [arXiv:1207.7235 [hep-ex]].

\bibitem{Farina:2013mla}
  M.~Farina, D.~Pappadopulo and A.~Strumia,
  JHEP {\bf 1308}, 022 (2013)
  [arXiv:1303.7244 [hep-ph]].

\bibitem{Bardeen:1995kv}
  W.~A.~Bardeen,
  FERMILAB-CONF-95-391-T.

\bibitem{Heikinheimo:2013fta}
  M.~Heikinheimo, A.~Racioppi, M.~Raidal, C.~Spethmann and K.~Tuominen,
  arXiv:1304.7006 [hep-ph].

\bibitem{related}
  R.~Hempfling,
  Phys.\ Lett.\ B {\bf 379}, 153 (1996)
  [hep-ph/9604278];
  K.~A.~Meissner and H.~Nicolai,
  Phys.\ Lett.\ B {\bf 648}, 312 (2007)
  [hep-th/0612165];
  W.~-F.~Chang, J.~N.~Ng and J.~M.~S.~Wu,
  Phys.\ Rev.\ D {\bf 75}, 115016 (2007)
  [hep-ph/0701254 [HEP-PH]];
  R.~Foot, A.~Kobakhidze, K.~.L.~McDonald and R.~.R.~Volkas,
  Phys.\ Rev.\ D {\bf 76}, 075014 (2007)
  [arXiv:0706.1829 [hep-ph]];
  R.~Foot, A.~Kobakhidze, K.~L.~McDonald and R.~R.~Volkas,
  Phys.\ Rev.\ D {\bf 77}, 035006 (2008)
  [arXiv:0709.2750 [hep-ph]];
  S.~Iso, N.~Okada and Y.~Orikasa,
  Phys.\ Lett.\ B {\bf 676}, 81 (2009)
  [arXiv:0902.4050 [hep-ph]];
  M.~Holthausen, M.~Lindner and M.~A.~Schmidt,
  Phys.\ Rev.\ D {\bf 82}, 055002 (2010)
  [arXiv:0911.0710 [hep-ph]];
  R.~Foot, A.~Kobakhidze and R.~R.~Volkas,
  Phys.\ Rev.\ D {\bf 82}, 035005 (2010)
  [arXiv:1006.0131 [hep-ph]]; 
  L.~Alexander-Nunneley and A.~Pilaftsis,
  JHEP {\bf 1009}, 021 (2010)
  [arXiv:1006.5916 [hep-ph]]; 
  T.~Hur and P.~Ko,
  Phys.\ Rev.\ Lett.\  {\bf 106}, 141802 (2011)
  [arXiv:1103.2571 [hep-ph]]; 
  S.~Iso and Y.~Orikasa,
  PTEP {\bf 2013}, 023B08 (2013)
  [arXiv:1210.2848 [hep-ph]];
  C.~Englert, J.~Jaeckel, V.~V.~Khoze and M.~Spannowsky,
  JHEP {\bf 1304}, 060 (2013)
  [arXiv:1301.4224 [hep-ph]]; 
  E.~J.~Chun, S.~Jung and H.~M.~Lee,
  Phys.\ Lett.\ B {\bf 725}, 158 (2013)
  [arXiv:1304.5815 [hep-ph]]; 
  T.~Hambye and A.~Strumia,
  Phys.\ Rev.\ D {\bf 88}, 055022 (2013)
  [arXiv:1306.2329 [hep-ph]];
  V.~V.~Khoze and G.~Ro,
  JHEP {\bf 1310}, 075 (2013)
  [arXiv:1307.3764]; 
  C.~D.~Carone and R.~Ramos,
  Phys.\ Rev.\ D {\bf 88}, 055020 (2013)
  [arXiv:1307.8428 [hep-ph]];
  G.~Marques Tavares, M.~Schmaltz and W.~Skiba,
  Phys.\ Rev.\ D {\bf 89}, 015009 (2014)
  [arXiv:1308.0025 [hep-ph]];
  A.~Farzinnia, H.~-J.~He and J.~Ren,
  Phys.\ Lett.\ B {\bf 727}, 141 (2013)
  [arXiv:1308.0295 [hep-ph]]; 
  O.~Antipin, M.~Mojaza and F.~Sannino,
  arXiv:1310.0957 [hep-ph]; 
  M.~Holthausen, J.~Kubo, K.~S.~Lim and M.~Lindner,
  JHEP {\bf 1312}, 076 (2013)
  [arXiv:1310.4423 [hep-ph]]; 
  S.~Abel and A.~Mariotti,
  arXiv:1312.5335 [hep-ph];
  C.~T.~Hill,
  Phys.\ Rev.\ D {\bf 89}, 073003 (2014)
  [arXiv:1401.4185 [hep-ph]]; 
  B.~Radovcic and S.~Benic,
  Phys.\ Lett.\ B {\bf 732}, 91 (2014)
  [arXiv:1401.8183 [hep-ph]];
  A.~de Gouvea, D.~Hernandez and T.~M.~P.~Tait,
  Phys.\ Rev.\ D {\bf 89}, 115005 (2014)
  [arXiv:1402.2658 [hep-ph]]; 
  J.~Kubo, K.~S.~Lim and M.~Lindner,
  arXiv:1403.4262 [hep-ph]; 
  V.~V.~Khoze, C.~McCabe and G.~Ro,
  arXiv:1403.4953 [hep-ph].


\bibitem{Hambye:2007vf}
  T.~Hambye and M.~H.~G.~Tytgat,
  Phys.\ Lett.\ B {\bf 659}, 651 (2008)
  [arXiv:0707.0633 [hep-ph]].

\bibitem{Meissner:2007xv} 
  K.~A.~Meissner and H.~Nicolai,
  Phys.\ Lett.\ B {\bf 660}, 260 (2008)
  [arXiv:0710.2840 [hep-th]];


\bibitem{Pilaftsis:1997jf} 
  A.~Pilaftsis,
  Phys.\ Rev.\ D {\bf 56}, 5431 (1997)
  [hep-ph/9707235].

\bibitem{OtherNeutrino}
For other attempts to reconcile the seesaw mechanism with large quadratic corrections to the Higss mass see 
  F.~Bazzocchi and M.~Fabbrichesi,
  Phys.\ Rev.\ D {\bf 87}, no. 3, 036001 (2013)
  [arXiv:1212.5065 [hep-ph]]; 
  M.~Fabbrichesi and S.~T.~Petcov,
  Eur.\ Phys.\ J.\ C {\bf 74}, 2774 (2014)
  [arXiv:1304.4001 [hep-ph]].

\bibitem{Deshpande:1977rw}
  N.~G.~Deshpande and E.~Ma,
  Phys.\ Rev.\ D {\bf 18}, 2574 (1978).

\bibitem{Ma:2006km}
  E.~Ma,
  Phys.\ Rev.\ Lett.\  {\bf 81}, 1171 (1998)
  [hep-ph/9805219]; 
  E.~Ma,
  Phys.\ Rev.\ D {\bf 73}, 077301 (2006)
  [hep-ph/0601225]

\bibitem{Coleman:1973jx} 
  S.~R.~Coleman and E.~J.~Weinberg,
  Phys.\ Rev.\ D {\bf 7}, 1888 (1973).

\bibitem{Carone:1993xc} 
  C.~D.~Carone and H.~Georgi,
  Phys.\ Rev.\ D {\bf 49}, 1427 (1994)
  [hep-ph/9308205].


\bibitem{Hambye:2009pw}
  T.~Hambye, F.~-S.~Ling, L.~Lopez Honorez and J.~Rocher,
  JHEP {\bf 0907}, 090 (2009)
  [Erratum-ibid.\  {\bf 1005}, 066 (2010)]
  [arXiv:0903.4010 [hep-ph]].

\bibitem{Chen:2009gd} 
  C.~-H.~Chen, C.~-Q.~Geng and D.~V.~Zhuridov,
  JCAP {\bf 0910}, 001 (2009)
  [arXiv:0906.1646 [hep-ph]]; 
  R.~Bouchand and A.~Merle,
  JHEP {\bf 1207} (2012) 084
  [arXiv:1205.0008 [hep-ph]].

\bibitem{Barbieri:2006dq}
  R.~Barbieri, L.~J.~Hall and V.~S.~Rychkov,
  Phys.\ Rev.\ D {\bf 74}, 015007 (2006)
  [hep-ph/0603188].


\bibitem{DMHiggs}
  J.~McDonald,
  Phys.\ Rev.\ D {\bf 50}, 3637 (1994)
  [hep-ph/0702143 [HEP-PH]]; 
  C.~P.~Burgess, M.~Pospelov and T.~ter Veldhuis,
  Nucl.\ Phys.\ B {\bf 619}, 709 (2001)
  [hep-ph/0011335];
  H.~Davoudiasl, R.~Kitano, T.~Li and H.~Murayama,
  Phys.\ Lett.\ B {\bf 609}, 117 (2005)
  [hep-ph/0405097]; 
  S.~Andreas, T.~Hambye and M.~H.~G.~Tytgat,
  JCAP {\bf 0810}, 034 (2008)
  [arXiv:0808.0255 [hep-ph]].


\bibitem{Akerib:2013tjd}
  D.~S.~Akerib {\it et al.}  [LUX Collaboration],
  arXiv:1310.8214 [astro-ph.CO].

\bibitem{Ade:2013zuv}
  P.~A.~R.~Ade {\it et al.}  [Planck Collaboration],
  arXiv:1303.5076 [astro-ph.CO].


\bibitem{Hill:1985tg}
  C.~T.~Hill, C.~N.~Leung and S.~Rao,
  Nucl.\ Phys.\ B {\bf 262}, 517 (1985).

\bibitem{Ade:2014xna}
  P.~A.~R.~Ade {\it et al.}  [BICEP2 Collaboration],
  arXiv:1403.3985 [astro-ph.CO].

\bibitem{Low:2014cba}
  M.~Low and L.~-T.~Wang,
  arXiv:1404.0682 [hep-ph].




\end{thebibliography}
\end{document}